# ASTROPHYSICAL AND STRUCTURAL PARAMETERS, AND DYNAMICAL EVOLUTION OF THE OPEN CLUSTERS NGC 1245 AND NGC 2099

Hikmet ÇAKMAK [1, *] , Orhan GÜNEŞ [2,] , Yüksel KARATAŞ [1]

[1] Department of Astronomy and Space Sciences, Faculty of Science, İstanbul University, İstanbul, Turkey
[2] İstanbul Medeniyet University, Faculty of Literature, Department of Science History, İstanbul, Turkey

## ABSTRACT

We derive astrophysical and structural parameters of open clusters NGC 1245 and NGC 2099 from 2MASS $JHK_s$ and Gaia DR2 photometric / astrometric data bases. Their likely members have been determined from Gaia DR2 proper motion data. Our E(B – V) values (2MASS) are slightly smaller than the literature values, whereas our E(B – V) values (Gaia DR2) agree with the literature within the uncertainties. Their distance moduli/distances and ages are in a good agreement with the early studies. NGC 1245 has steep negative core mass function slope (MFs) ($\chi_{core} = -1.24$). Its halo ($\chi_{halo} = +0.78$) and overall ($\chi_{overall} = -0.95$) MFs indicate small-scale mass segregation from its core to the outer regions, due to its $[t_{rlx}(overall), \tau_{overall}] = [302\ Myr, 5]$. The MFs of NGC 2099 is very negative steep ($\chi_{core} = -2.67$) in the core, and quite positive steep ($\chi_{halo} = +1.41$) in the halo. This kind of MF slope steeping from the core to the outskirts indicates that low-mass stars in the core are transferred to the cluster's outskirts, while massive stars sink in the core, because of mass segregation. NGC 2099's flat overall MFs ($\chi_{overall} = +0.91$) and its $\tau_{overall} = 8$ indicate mass segregation. These OCs with the relatively large masses 8700 $M_\odot$ (NGC 1245) and 5660 $M_\odot$ (NGC 2099), which locate at $R_{GC} > 9$ kpc, expose to external perturbations such as tidal effects and shock waves, and they lose their stars in low-proportions.

## 1. INTRODUCTION

Astrophysical parameters (reddening, distance, age), structural parameters (core and cluster radii), core and overall masses, mass function slopes (MFs), relaxation time and evolutionary parameters are needed for the interpretation of the dynamical evolution of the Galactic open clusters (OCs). The stars inside the OCs undergo internal and external perturbations such as stellar evolution, mass segregation and evaporation, tidal interactions with Galactic disk and bulge, and collisions with Giant Molecular Clouds (GMCs) (Lamers and Gieles, 2006[1]), As clusters age, these mechanisms accelerate the internal dynamical evolution, which leads to important changes in the structure. All these processes produce a mass loss that may lead to the cluster dissolution into the field. Mass segregation drives the low mass stars to the outer parts of the clusters. A steep MFs of the outer skirts of the clusters presents a sign of mass segregation (Ann and Lee, 2002[2]).

The OCs are located in Galactic disk, so background contamination is large. This makes it quite difficult to select the cluster stars from the background. For this respective, Gaia DR2 astrometric data (proper motions and parallaxes) give us a chance to separate cluster stars from field stars. 2MASS JH Ks and Gaia DR2 $G$ - ($G_{BP} - G_{RP}$) photometries give us a good opportunity to obtain the astrophysical parameters for a uniform and homogeneous analysis.

The metal-rich and old open cluster NGC 1245 is located in the Perseus arm. NGC 2099 which is the focus of many studies is a relatively rich and large star cluster. Both OCs locate at second Galactic quadrant (Figure 1). In this paper, we shed light on the problems mentioned above with the recent Gaia DR2 astrometric / photometric data combined with JH Ks photometry of the two OCs.

This paper is organized as follows. An introduction is presented in Section 1. 2MASS *JHKs* and Gaia DR2 *G* - ($G_{BP}$ – $G_{RP}$) photometries, cluster membership technique, determination of cluster dimensions and structural parameters are given in Section 2. The astrophysical parameters such as reddenings, distance moduli/distances and ages are presented in Section 3. The derivation of mass and MFs, relaxation and evolutionary time parameters of the two OCs are given in Sections 4-5. A discussion and conclusion is given in Section 6 together with the comparison to the literature and the dynamical evolution.

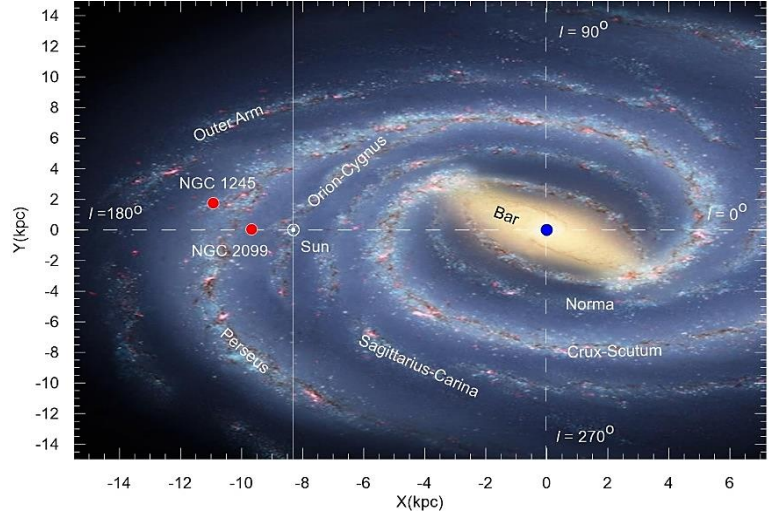

**Figure 1.** Spatial distributions of NGC 1245 and NGC 2099 (red filled circles). The Sun shown with white circle is at a distance 8.2 ± 0.1 kpc (Bland-Hawthorn and Gerhard, 2016[3]) from galactic center (blue filled circle). Background image[1] is adapted from the image credit by Robert Hurt, IPAC; Bill Saxton, NRAO/AUI/NSF.

## 2. SEPARATION OF CLUSTER MEMBERS AND THEIR DIMENSIONS

For NGC 1245 and NGC 2099, the 2MASS *JHKs* photometry[2] (Skrutskie et al. 2006[4]) has been extracted from VizieR[3]. Their Gaia DR2 astrometric ($\mu_\alpha, \mu_\delta$ and $\varpi$) and photometric data (*G* - ($G_{BP}$ – $G_{RP}$)) are taken from the CDS X-Match Service[4] of University de Strasbourg/CNRS. We extracted all the Gaia DR2 sources within a radius of 15 arcmin from the cluster centers of our targets (Table 1). To make a match, we set 1 arcsec as the maximum difference in $\alpha_{2000}$ and $\delta_{2000}$ for both datasets. Their finding charts from *Aladin Sky Atlas*[5] are displayed in Figure 2 to show their overall appearance in the sky. Their equatorial and Galactic coordinates are listed in Table 1.

**Table 1.** The equatorial and Galactic coordinates of NGC 1245 and NGC 2099, taken from WEBDA.

| Parameter | NGC 1245 | NGC 2099 |
|---|---|---|
| $\alpha_{2000}$ (h m s) | 03 14 53 | 05 52 18 |
| $\delta_{2000}$ (° ′ ″) | +47 14 00 | +32 33 12 |
| $\ell$ (°) | 146.68 | 177.63 |
| $b$ (°) | –8.92 | +3.09 |

---

[1] https://www.universetoday.com/102616/our-place-in-the-galactic-neighborhood-just-got-an-upgrad

[2] The Two Micron All Sky Survey, available at http://www.ipac.caltech.edu/2mass/releases/allsky/

[3] http://vizier.u-strasbg.fr/viz-bin/VizieR?-source=II/246.

[4] http://cdsxmatch.u-strasbg.fr/xmatch

[5] https://aladin.u-strasbg.fr/AladinLite/

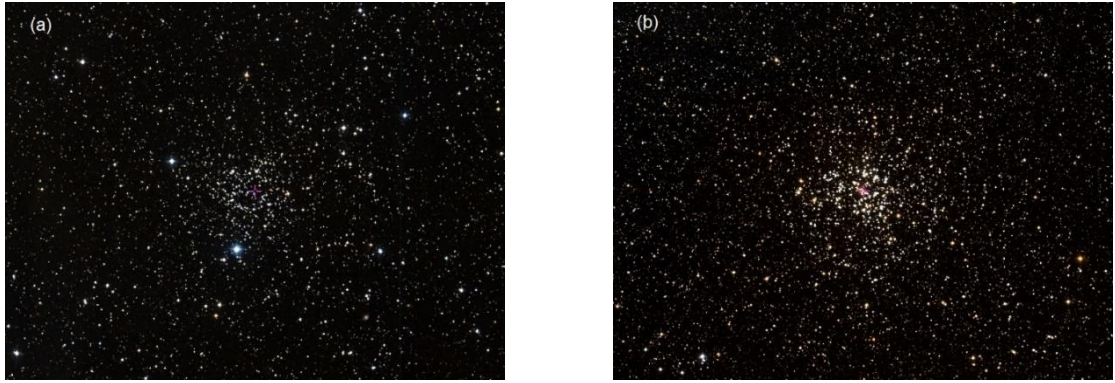

**Figure 2.** Finding charts of NGC 1245 (a) and NGC 2099 (b). Red plus symbols show the central coordinates.

Their stellar radial density profiles (RDPs) have been built from Gaia DR2 photometry (Figure 3). These RDPs are built by counting stars between concentric rings of increasing width with distance to its center. The external ring width of the stellar comparison field was selected as 15 arc-minute. The number and width of rings were optimized so that the resulting RDP had adequate spatial resolution with moderate 1σ Poisson errors, as emphasized by Bonatto and Bica (2007)[5]. The solid curves in Figure 3 represent the fitted King's profile (King, 1966[6]) obtained by using the formula

$$\sigma(R) = \sigma_{bg} + \frac{\sigma_0}{1 + \left(\frac{R}{R_c}\right)^2} \tag{1}$$

where $R$ is the distance to cluster center, $\sigma_{bg}$ is the residual background density, $\sigma_0$ is the central density of stars, and $R_c$ is the core radius. The horizontal gray bar shows the stellar background level measured in the comparison field, and the 1σ profile fit uncertainty is shown by red shaded region. The core and cluster radii of each cluster determined by curve fitting are obtained as ($R_{core}$, $R_{RDP}$) = (3.12' ± 0.72, 22.51' ± 1.45) for NGC 1245, ($R_{core}$, $R_{RDP}$) = (5.48' ± 0.76, 32.58' ± 1.44) for NGC 2099, respectively. These values transform to ($R_{core}$, $R_{RDP}$)=(2.89 ± 0.67, 20.97 ± 1.95) pc for NGC 1245 and (2.20 ± 0.30, 13.07 ± 0.58) pc for NGC 2099. Our values are in good concordance with the literature within the uncertainties (See Cols.2-3 of Table 6).

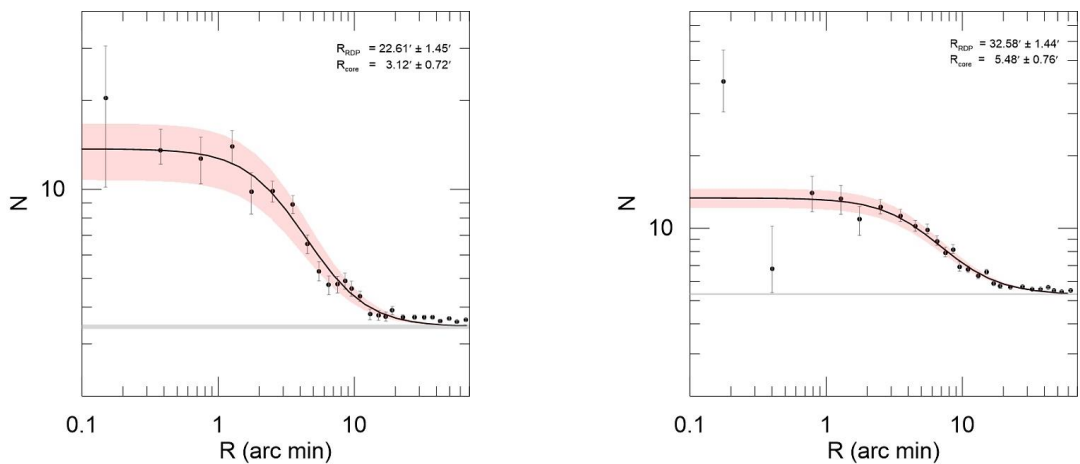

**Figure 3.** Stellar RDPs of NGC 1245 (*Left*) and NGC 2099 (*Right*) are shown as filled dots with their errors. Solid line indicates the best-fit King profile and horizontal gray bar displays the stellar background level measured in the comparison field. Red shaded region represents the 1σ King fit uncertainty. The core and cluster radii of each cluster as arc min are shown on the top right corner.

The $\mu_\alpha$ versus $\mu_\delta$ (*mas yr$^{-1}$*) Vector Point Diagram, (VPD) for all stars of the two OCs (filled dots) is shown in top panels of Figure 4 along with ($G$, $G_{BP} - G_{RP}$) plots (bottom panels). Grey dots in the bottom panels denote the field stars inside R = 15.0 arcmin centered on the two OCs. The frequency distributions of $\mu_\alpha$ and $\mu_\delta$ are shown on the top and right sides of VPDs in the top panels. Note that two distinct populations are clearly visible; cluster stars are tightly distributed while field stars are more scattered. For NGC 2099 (right panel), it appears that there are two concentrations with two peaks in proper motion components. Cluster population seems to be concentrated at ($\mu_\alpha$, $\mu_\delta$) ~ (+1.90, -5.5) mas yr$^{-1}$.

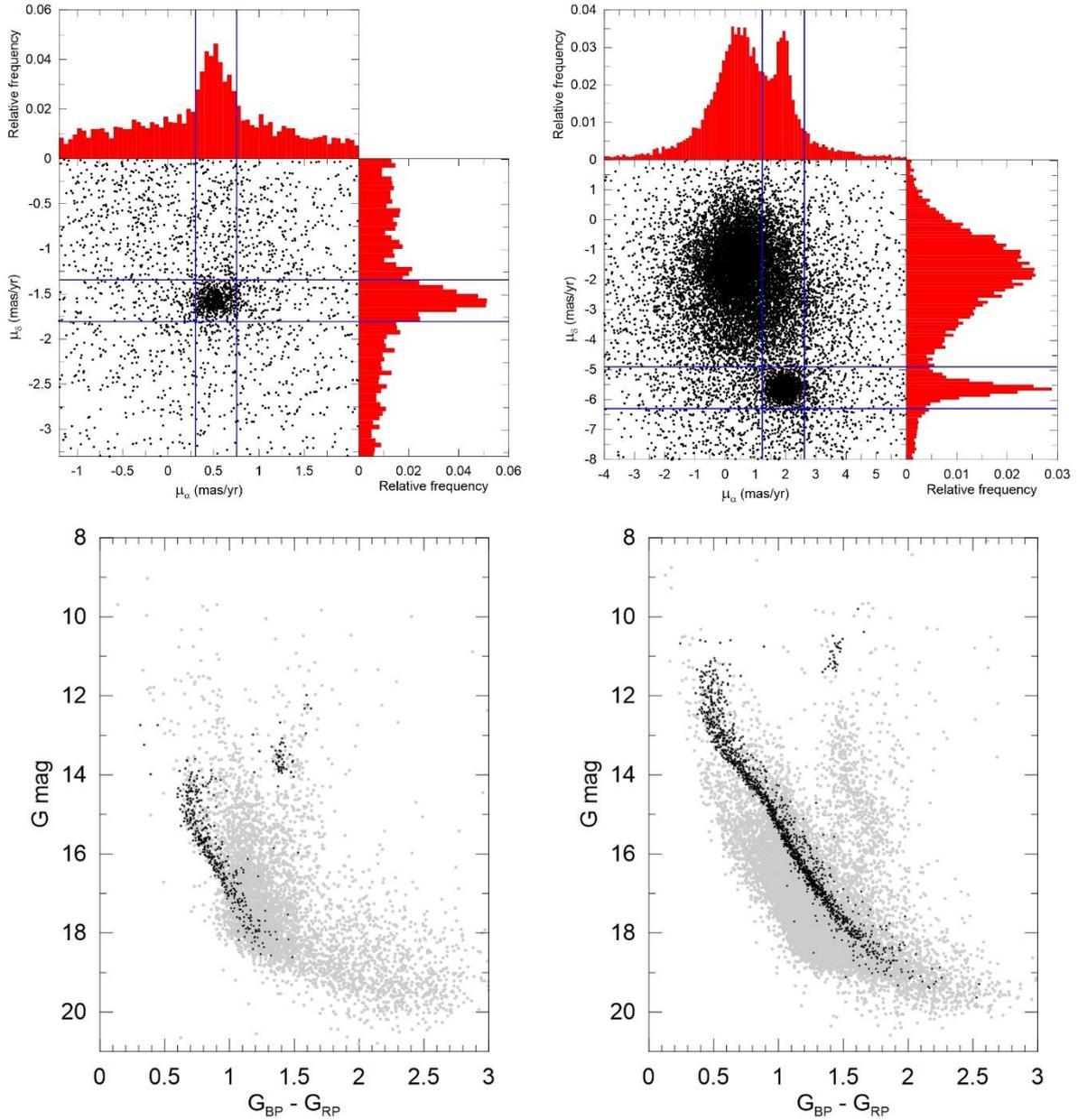

**Figure 4.** The $\mu_\alpha$ versus $\mu_\delta$ for NGC 1245 (577 black dots, *left panel*) and NGC 2099 (1926 black dots, *right panel*). The field stars inside 15 arcmin are shown with grey dots. The fitted proper motion circles (red ones in upper panels) denote the radii of 0.23 mas yr$^{-1}$ for NGC 1245 and 0.65 mas yr$^{-1}$ for NGC 2099, which are considered as the likely members. Single stellar cluster sequences of the probable members (filled dots of bottom panels) are clearly visible on ($G$, $G_{BP} - G_{RP}$).

The proper motion radii (shown with red circles on upper panels in Figure 4) of 0.23 mas yr $^{-1}$ for NGC 1245 and 0.65 mas yr $^{-1}$ for NGC 2099 around the centers of the VPDs define the membership criteria. These proper motion radii have been empirically fitted until the likely members inside these radii provide good single stellar sequence on ($G$, $G_{BP} - G_{RP}$) plots (bottom panels of Figure 4). The number of the probable cluster members inside the proper motion radii is 577 for NGC 2145 and 1926 for NGC 2099, respectively. These proper motion radii are determined via the mathematical equations, $x = x_0 + r\ Cos(\theta)$ and $y = y_0 + r\ Sin(\theta)$. Here, ($x_0$, $y_0$) are the median values of ($\mu_\alpha$, $\mu_\delta$) (mas yr $^{-1}$), the radius $r = \sqrt{\sigma_{\mu\alpha}^2 + \sigma_{\mu\delta}^2}$ mas yr $^{-1}$, and $\theta = 0°$ to $360°$.

The median proper motion components of the probable members are ($\mu_\alpha$, $\mu_\delta$) = (0.493 ± 0.098, –1.578 ± 0.080) mas yr $^{-1}$ for NGC 1245 and ($\mu_\alpha$, $\mu_\delta$) = (1.929 ± 0.203, –5.634 ± 0.185) mas yr $^{-1}$ for NGC 2099, respectively. These median values are in good agreement with ($\mu_\alpha$, $\mu_\delta$) = (0.504 ± 0.113, –1.579 ± 0.10) mas yr $^{-1}$ for NGC 1245 ($N = 589$) and ($\mu_\alpha$, $\mu_\delta$) = (1.924 ± 0.201, –5.648 ± 0.175) mas yr $^{-1}$ for NGC 2099 ($N = 1710$), respectively, given by Cantat-Gaudin et al. (2018)[7]. Having applied these proper motion criteria to our OCs, the likely cluster members are now appropriate for determining the astrophysical parameters.

The uncertainties of ($\sigma_{\mu\alpha}$, $\sigma_{\mu\delta}$, $\sigma_\varpi$) are (0.134, 0.109, 0.076) for NGC 1245 and (0.819, 0.673, 0.167) for NGC 2099. These limits nearly remain within the uncertainties of ($\sigma_{\mu\alpha}$, $\sigma_{\mu\delta}$, $\sigma_\varpi$) = (0.28, 0.24, 0.16) for $G < 18$ mag., reported by the Gaia Collaboration, Lindegren et al. (2018)[8] (see their table B.1). For the likely members with $\sigma_\varpi/\varpi < 0.20$ (Figure 5), the median value of Gaia DR2 parallaxes is 0.294 ± 0.041 mas (N = 284) for NGC 1245 and 0.670 ± 0.065 mas (N = 1356) for NGC 2099. Hence these values correspond to 3.4 ± 0.5 kpc (NGC 1245) and 1.5 ± 0.1 kpc (NGC 2099).

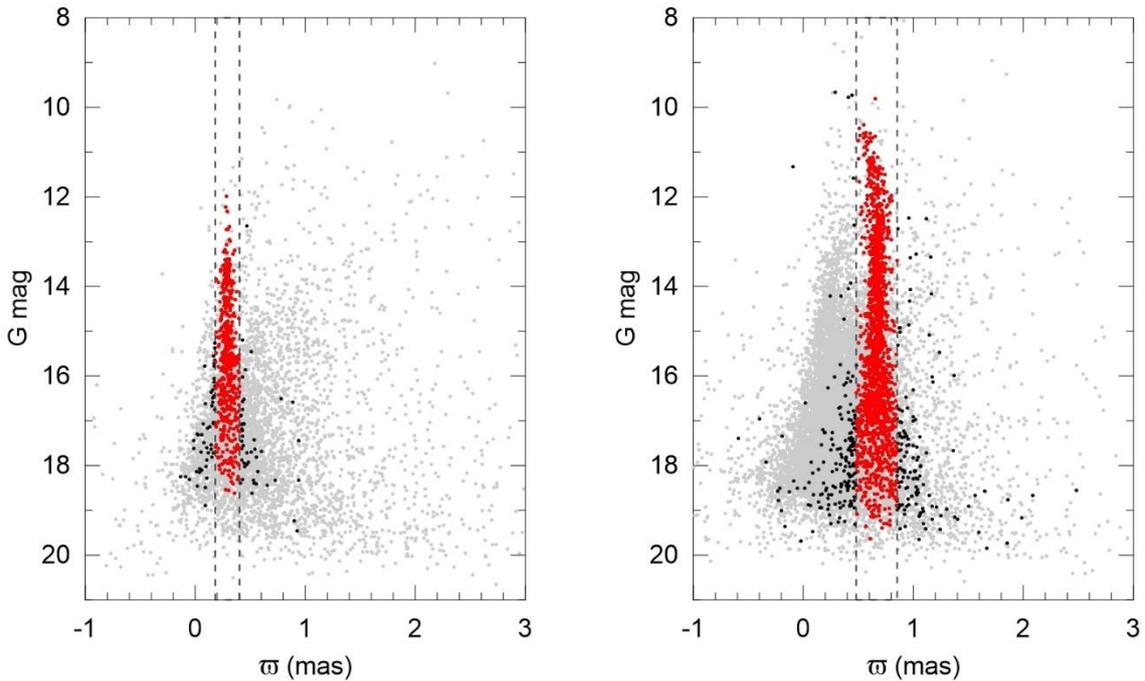

**Figure 5.** *G mag* versus $\varpi$ (mas) plots for NGC 1245 (*left panel*) and NGC 2099 (*right panel*). The red dots represent the cluster members within the 1σ uncertainty.

## 3. ASTROPHYSICAL PARAMETERS OF NGC 1245 AND NGC 2099

The (*J*, *J* – *H*) and (*G*, $G_{BP}$ – $G_{RP}$) color magnitude diagrams (CMDs) for the probable cluster members of NGC 1245 and NGC 2099 are displayed in left and right panels of Figure 6 – 7. The filters (shown in Figure 6) are wide enough to reflect the color distributions of main sequence and evolved stars within 1σ photometric uncertainty. The distribution of cluster members in (*G*, $G_{BP}$ – $G_{RP}$) appears to be tighter than the distribution in (*J*, *J* – *H*). Jacobson et al. (2011)[9] give the spectroscopic metal abundance of NGC 1245 as [Fe/H] = –0.04 ± 0.04. For NGC 2099, a spectroscopic metal abundance, [Fe/H] = +0.01 ± 0.05 is given by Pancino et al. (2010)[10]. These values convert to the heavy element abundance, Z = 0.015 (solar abundance). Therefore, Padova isochrones of Bressan et al. (2012)[11] (hereafter referred as B12) of Z = +0.015 are fitted to the (*J*, *J* – *H*) and (*G*, $G_{BP}$ – $G_{RP}$) CMDs, and thus the reddenings, distance moduli and ages are obtained. As seen from Figure 6 – 7, the B12 isochrones fit well the main sequence (MS), turn-off (TO) and Red Giant/Red Clump (RG/RC) regions on the CMDs. For (*J*, *J* – *H*), B12 isochrones gave us 1.50 Gyr for NGC 1245 and 0.90 Gyr for NGC 2099, respectively. From the fitting the B12 isochrone to (*G*, $G_{BP}$ – $G_{RP}$), we obtained an age of 1.50 Gyr for NGC 1245 and 0.80 Gyr for NGC 2099, respectively.

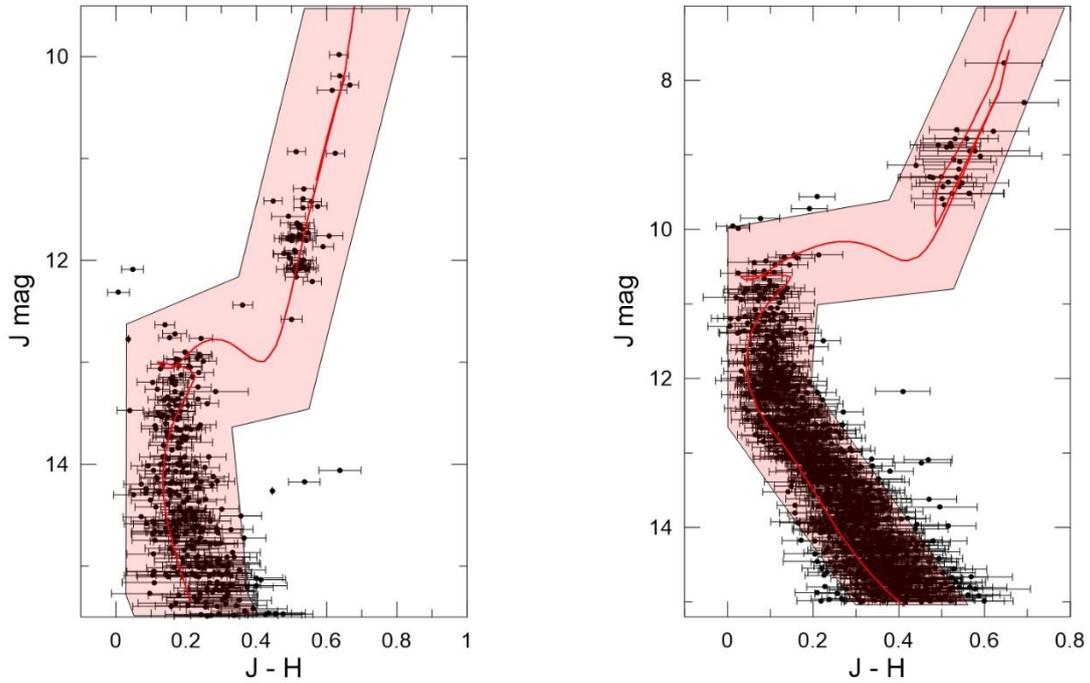

**Figure 6.** *J* × (*J* – *H*) CMDs for the members of NGC 1245 (*left panel*) and NGC 2099 (*right panel*). The solid red lines represent the fitted 1.50 Gyr (NGC 1245) and 0.90 Gyr (NGC 2099) B12 isochrones for Z = +0.015 (solar) abundance. The CMD filter used to isolate cluster MS/evolved stars is shown with the red shaded area.

The obtained reddenings in *E*(*J* – *H*) are converted to *E*(*B* – *V*) by considering the relations $A_J / A_V$ = 0.276, $A_H / A_V$ =0.176, $A_{Ks} / A_V$ = 0.118, $A_J$ =2.76 × *E*(*J* – *H*), and *E*(*J* – *H*) = 0.33 × *E*(*B* – *V*) (Dutra et al. 2002[12]). For the reddenings in *E*($G_{BP}$ – $G_{RP}$), we adopt *E*(*B* – *V*) = 0.775 × *E*($G_{BP}$ – $G_{RP}$), (Bragaglia et al. 2018[13]). Their resulting *E*(*J* – *H*), *E*($G_{BP}$ – $G_{RP}$), *E*(*B* – *V*), distance moduli, distances and ages are given in Cols. 3 – 8 of Table 2. The Galactocentric distances are estimated as $R_{GC}$ = 10.99 kpc for NGC 1245 and $R_{GC}$ = 9.59 kpc for NGC 2099, respectively. Here we adopt $R_⊙$ =8.2 ± 0.1 kpc of Bland-Hawthorn and Gerhard (2016)[3].

The uncertainties of *E*(*J* – *H*) (Table 2) have been estimated moving the B12 isochrones up and down, back and forward and in direction of the reddening vector on (*J*, *J* – *H*) and (*G*, $G_{BP}$ – $G_{RP}$). The uncertainties of distance moduli and ages (Table 2) are due to both the photometric errors and the fitting the appropriate isochrone to the stars on both CMDs. Güneş et al. (2017)[14] discuss that *JHKs* photometry is unsensitive to metallicity. Metallicity affects significantly the distance and the age of a cluster, i.e. the less Z is, the shorter the distance and larger the age. As stated by Bonatto and Bica (2009)[15], any metallicity for the range of +0.006 < Z < +0.015 produces acceptable solutions for the astrophysical parameters, due to the filters of 2MASS. Kharchenko et al. (2013)[16] discuss that the metallicities of OCs are of the order of solar value, and the Padova isochrones of the same age and different Z mildly affect the derived cluster parameters.

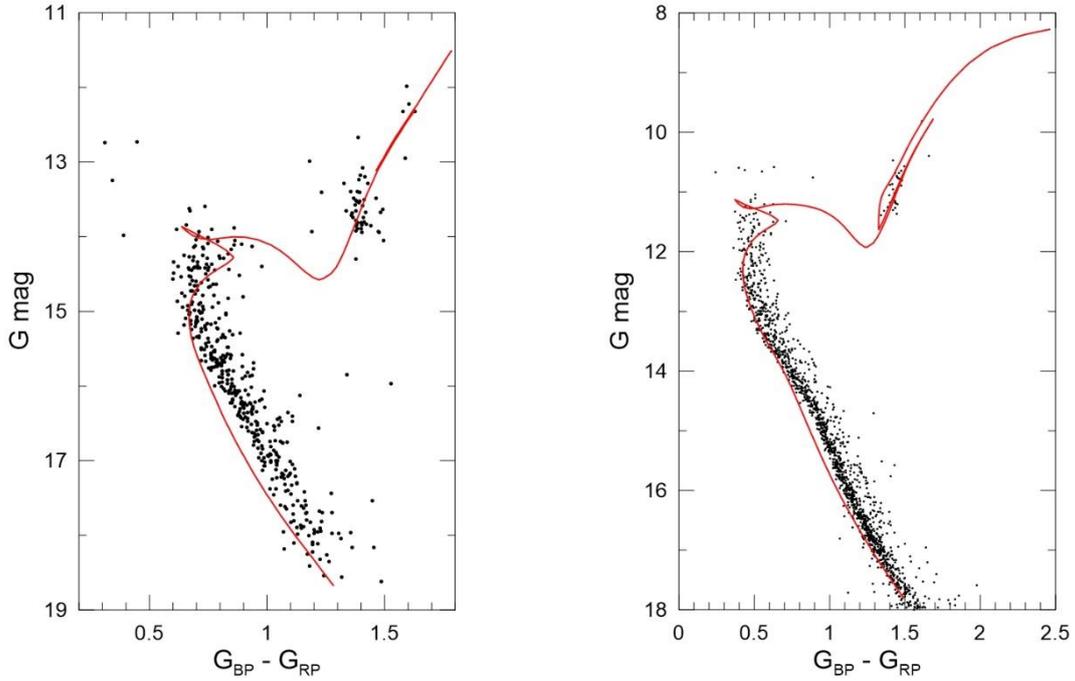

**Figure 7.** *(G, $G_{BP} - G_{RP}$)* diagrams for NGC 1245 (*left*) and NGC 2099 (*right*). Red curve denotes B12 isochrone for Z = +0.015.

**Table 2.** The obtained astrophysical parameters of our targets for two color indices.

| Cluster | Color | $E(\lambda_1 - \lambda_2)$ | $E(B - V)$ | $(V_0 - M_V)$ | d (kpc) | log(A) | A (Gyr) |
|---|---|---|---|---|---|---|---|
| NGC 1245 | $(J - H)$ | 0.01 ± 0.04 | 0.030 ± 0.12 | 12.52 ± 0.08 | 3.19 ± 0.11 | 9.18 ± 0.10 | 1.50 ± 0.35 |
|  | $(G_{BP} - G_{RP})$ | 0.24 ± 0.06 | 0.186 ± 0.05 | 12.41 ± 0.08 | 3.03 ± 0.11 | 9.18 ± 0.06 | 1.50 ± 0.20 |
| NGC 2099 | $(J - H)$ | 0.003 ± 0.018 | 0.010 ± 0.05 | 10.70 ± 0.08 | 1.38 ± 0.04 | 8.95 ± 0.05 | 0.90 ± 0.10 |
|  | $(G_{BP} - G_{RP})$ | 0.23 ± 0.06 | 0.178 ± 0.05 | 10.63 ± 0.08 | 1.33 ± 0.05 | 8.90 ± 0.05 | 0.80 ± 0.09 |

## 4. MASS AND MASS FUNCTION SLOPE

We have utilized 2MASS *JHKs* data for determining the masses of the likely cluster members of NGC 1245 and NGC 2099. From the procedures of Bonatto and Bica (2005)[17] and Bica et al. (2006)[18], we have derived the stellar masses of the cluster members from their mass functions (MFs) for the observed main sequence (MS) mass range. Their luminosity functions from the (J, J – H) diagram have been converted into mass function slopes (MFs) via the mass-luminosity relations of B12 isochrones for the ages (Table 2). The relations of $\Phi(m)$(*stars* $m_\odot^{-1}$) versus $m_\odot$ are displayed in Figure 8 in terms of their cluster regions.

Having fitted the MS MFs of the two OCs to function $\Phi(m) \propto m^{-(1+\chi)}$, the MF slopes ($\chi$) have been determined for the different regions (core, halo and overall). The number of stars as MS and evolved star, MF slopes ($\chi$), and masses extrapolated to 0.08 $m_\odot$ for different regions of the two OCs are given in Table 3. The lower MS is not accessed on the ($J, J – H$) diagrams of the two OCs. Therefore, we considered Kroupa's MF[6] to estimate the total stellar mass, down to the H-burning mass limit (0.08 $m_\odot$).

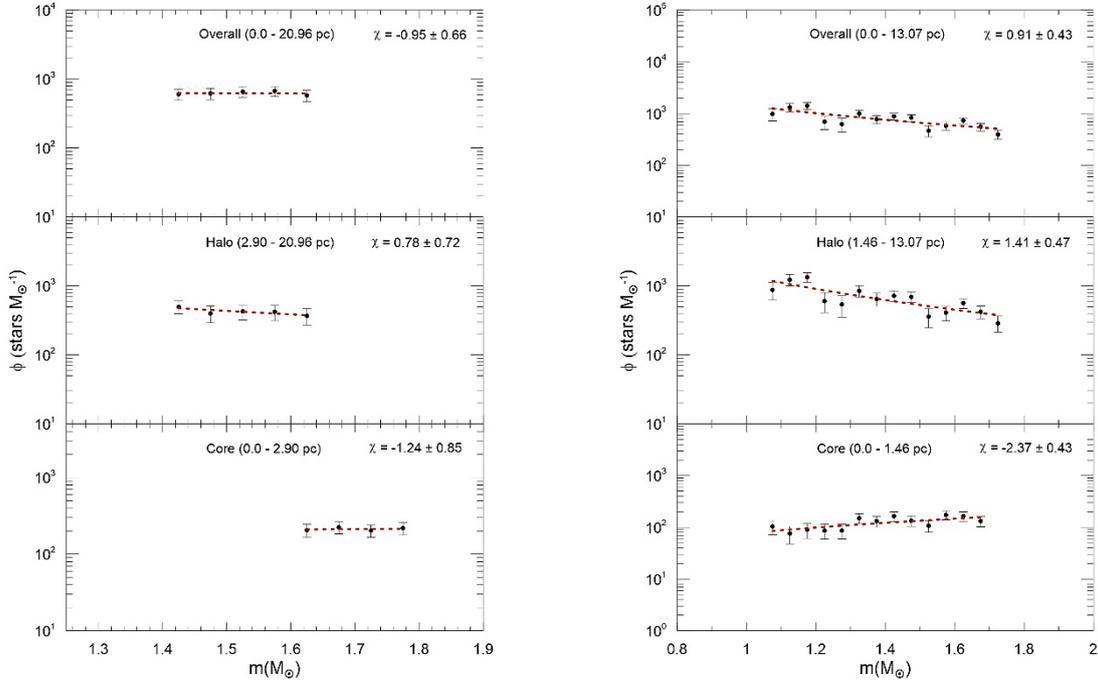

**Figure 8.** $\Phi(m)$(*stars $M_\odot^{-1}$*) versus $m(M_\odot)$ of NGC 1245 (*left*) and NGC 2099 (*right*), as a function of distance from the core.

## 5. RELAXATION TIME AND EVOLUTIONARY PARAMETER

To study the dynamical evolution from the parameters such as age, MFs, the location, cluster dimensions, the relaxation times and evolutionary parameters of the two OCs, we have estimated the relaxation times, $t_{rlx}$ from a relation, $t_{rlx} \approx 0.04 \left(\frac{N}{\ln N}\right)\left(\frac{R}{1pc}\right)$. Here $N$ is the number of stars inside the cluster radius, $R_{RDP}$. As an indicator of dynamical evolution, the evolutionary parameters are estimated from the relation of $\tau = Age/t_{rlx}$ (Bonatto and Bica, 2005[17]; Bonatto and Bica, 2006[19]; Bonatto and Bica, 2007[5]). The relaxation time $t_{rlx}$ is defined as the characteristic time-scale for a cluster to reach some level of energy equipartition, and the evolutionary parameter $\tau = Age/t_{rlx}$ is a good indicator of dynamical state. As emphasized by Bonatto and Bica (2006)[19], the two parameters present a signature that low-mass stars originally in the core are transferred to the cluster's outskirts, while massive stars sink in the core, which is known as mass-segregation. By adopting a typical $\sigma_v \approx 3$ *km s*$^{-1}$ (Binney and Merrifield, 1998[20]), the relaxation times [$t_{rlx}$(*overall*), $t_{rlx}$(*core*)] and the evolutionary parameters [$\tau_{overall}$, $\tau_{core}$)] for our sample OCs are estimated and listed in Table 4.

From propagating the errors in Age (Table 2), Radii (Table 6) and N (Table 3) into $t_{rlx}$ and $\tau$, the uncertainties of the evolutionary parameters ($\tau$) of the two OCs are estimated. This is responsible for a large uncertainty in $t_{rlx}$ (Table 4) and, consequently, a large uncertainty in the evolutionary parameter. Here, $t_{rlx}$ and $\tau$ are considered simply as an order of magnitude estimate.

---

[6] $\chi = 0.3 \pm 0.5$ Kroupa (2001) [39] for $0.08 < m_\odot < 0.5$, $\chi = 1.3 \pm 0.3$ for $0.5 < m_\odot < 1.0$, and $\chi = 1.3 \pm 0.7$ for $1.0 < m_\odot$.

**Table 3.** The number of stars, mass information, mass function slope, mass density, for cluster regions of our sample OCs for the cases of Evolved, Observed + Evolved, and Extrapolated + Evolved.

| Region (pc) | Evolved | | | Observed + Evolved | | Extrapolated + Evolved | |
|---|---|---|---|---|---|---|---|
| | $N^*$ (Stars) | $M_{evol}$ ($10^1 M_\odot$) | $\chi$ | $N^*$ ($10^2$ Stars) | $M_{obs}$ ($10^2 M_\odot$) | $N^*$ ($10^2$ Stars) | $M_{tot}$ ($10^2 M_\odot$) |
| (1) | (2) | (3) | (4) | (5) | (6) | (7) | (8) |
| NGC 1245 | | | $M = 1.43 - 1.78 M_\odot$ | | | | |
| 0.00 - 2.90 | 44 ± 7 | 8.1 ± 1.3 | -1.24 ± 0.85 | 1.21 ± 0.09 | 2.12 ± 0.79 | 7.81 ± 14.60 | 8.16 ± 5.06 |
| 2.90 - 20.96 | 70 ± 26 | 12.8 ± 4.8 | 0.78 ± 0.72 | 2.19 ± 0.30 | 3.76 ± 1.92 | 209 ± 200 | 72.10 ± 45.30 |
| 0.00 - 20.96 | 114 ± 27 | 21.0 ± 5.0 | -0.95 ± 0.66 | 3.40 ± 0.31 | 5.88 ± 2.29 | 245 ± 222 | 87.10 ± 46.40 |
| NGC 2099 | | | $M = 1.08 - 1.73 M_\odot$ | | | | |
| 0.00 - 1.46 | 21 ± 5 | 4.8 ± 1.0 | -2.67 ± 0.47 | 1.71 ± 0.09 | 2.57 ± 0.48 | 2.26 ± 0.31 | 2.98 ± 0.50 |
| 1.46 - 13.07 | 17 ± 11 | 3.7 ± 2.4 | 1.41 ± 0.47 | 7.79 ± 0.42 | 10.60 ± 2.07 | 183 ± 139 | 61.80 ± 25.90 |
| 0.00 - 13.07 | 38 ± 12 | 8.5 ± 2.6 | 0.91 ± 0.43 | 9.49 ± 0.44 | 13.20 ± 2.34 | 152 ± 113 | 56.60 ± 21.40 |

**Table 4.** Relaxation times and evolutionary parameters of the core and overall regions of the two OCs.

| Cluster | Core | | Overall | |
|---|---|---|---|---|
| | $t_{relax}$ (Myr) | $\tau$ | $t_{relax}$ (Myr) | $\tau$ |
| (1) | (2) | (3) | (4) | (5) |
| NGC 1245 | 14.6 ± 23.5 | 103 ± 165 | 302 ± 257 | 5.0 ± 4.3 |
| NGC 2099 | 5.2 ± 1.3 | 288 ± 83 | 197 ± 139 | 8.0 ± 5.0 |

## 6. DISCUSSION AND CONCLUSION

For NGC 1245, the reddenings are derived as $E(B – V) = 0.03 ± 0.12$ ($J, J – H$) and $E(B – V) = 0.19 ± 0.05$ ($G, G_{BP} – G_{RP}$). For NGC 2099, we obtained the reddenings as $E(B – V) = 0.01 ± 0.05$ ($J, J – H$) and $E(B – V) = 0.18 ± 0.05$ ($G, G_{BP} – G_{RP}$). Our reddenings of 2MASS are smaller than the other photometries (Table 5). For NGC 1245, the differences with the literature are at a levels of $\Delta E(B – V) = 0.06 – 0.25$ (2MASS) and $\Delta E(B – V) = 0.02 – 0.10$ (Gaia DR2). Discrepancies with the literature for NGC 2099 are $\Delta E(B – V) = 0.10 – 0.29$ (2MASS) and $\Delta E(B – V) = 0.03 – 0.12$ for (Gaia DR2), respectively. Our $E(B – V)$ values from ($G, G_{BP} – G_{RP}$) for both OCs are in good agreement with the literature values within the uncertainty limits.

For NGC 1245, the differences of the distances to the literature are at the level of 142 – 690 pc for 2MASS and 18 – 530 pc for Gaia DR2. These discrepancies for NGC 2099 fall in the range of 20 – 153 pc for 2MASS and 12 – 184 pc for Gaia DR2, respectively. Within the uncertainties, our distances are in good concordance with the literature values for 2MASS and Gaia DR2 Photometries. For the members with $\sigma_\varpi/\varpi < 0.20$, the median Gaia DR2 parallaxes of both OCs estimate the distances as 3400 ± 500 pc for NGC 1245 and 1500 ± 100 pc for NGC. 2099, and they are in good agreement with the ones of Cantat-Gaudin et al. (2018)[7] (Table 5).

The ages of NGC 1245 and NGC 2099 are 1.50 and 0.90 Gyr, respectively. Their ages are somewhat older than the ones of literature (Table 5). Note that both OCs have solar abundance (Z = 0.015).

**Table 5.** Comparing the astrophysical parameters of two OCs with the literature.

| Cluster | E(B-V) | (Vo - Mx) | d (pc) | Z | log Age | Age (Gyr) | Isochrone | Photometry | Ref. |
|---|---|---|---|---|---|---|---|---|---|
| NGC 1245 | 0.03 ± 0.12 | 12.52 ± 0.08 | 3190 ± 110 | 0.015 | 9.18 ± 0.10 | 1.50 ± 0.35 | Bressan et al. (2012)[11] | 2MASS | This paper |
|  | 0.19 ± 0.05 | 12.41 ± 0.08 | 3030 ± 110 | 0.015 | 9.18 ± 0.06 | 1.50 ± 0.20 | Bressan et al. (2012)[11] | GAIA | This paper |
|  | 0.09 | 12.22 | 2679 ± 156 | 0.019 | 9.05 | 1.12 | Girardi et al. (2002) [21] | 2MASS | Bukowiecki et al. (2011)[22] |
|  | 0.21 ± 0.01 | 12.27 ± 0.02 | 2800 ± 200 | 0.013 | 9.02 ± 0.01 | 1.04 ± 0.02 | Yi et al. (2001)[23] | CCD BVI | Burke et al. (2004)[24] |
|  | 0.26 | 12.42 | 3048 | 0.021 | 8.9 | 0.8 | Bressan et al. (1993)[25] | CCD BV | Carraro and Patat (1994)[26] |
|  | 0.28 ± 0.03 | 12.00 ± 0.15 | 2500 ± 200 | 0.014 | 9.04 ± 0.04 | 1.1 ± 0.1 | Bertelli et al. (1994)[27] | CCD CMT1T2 | Wee et al. (1996)[28] |
|  | 0.24 ± 0.05 | 12.25 ± 0.12 | 2818 ± 156 | 0.012 | 9.03 ± 0.04 | 1.08 ± 0.09 | Girardi et al. (2000)[29] | CCD VI | Lee et al. (2012)[30] |
|  | 0.29 ± 0.05 | 12.40 ± 0.30 | 3020 ± 418 | 0.014 | 8.95 ± 0.05 | 0.89 ± 0.10 | Girardi et al. (2000)[29] | CCD BV | Subramaniam (2003)[31] |
|  |  |  | 3460 ± 600 |  |  |  |  | Gaia DR2 parallax | Cantat-Gaudin et al. (2018)[7] |
|  |  |  | (w = 0.289 ± 0.05 mas) |  |  |  |  |  |  |
| NGC 2099 | 0.01 ± 0.05 | 10.70 ± 0.08 | 1380 ± 40 | 0.015 | 8.95 ± 0.05 | 0.90 ± 0.10 | Bressan et al. (2012)[11] | 2MASS | This paper |
|  | 0.18 ± 0.05 | 10.63 ± 0.08 | 1330 ± 50 | 0.015 | 8.90 ± 0.05 | 0.80 ± 0.09 | Bressan et al. (2012)[11] | GAIA | This paper |
|  | 0.11 | 10.54 | 1227 ± 72 | 0.019 | 8.9 | 0.79 | Girardi et al. (2002) [21] | 2MASS | Bukowiecki et al. (2011)[22] |
|  | 0.21 ± 0.03 | 10.90 ± 0.16 | 1514 ± 112 | 0.02 | 8.72 | 0.52 | Ventura et al. (1998) [32] | CCD CFHT | Kalirai et al. (2001)[33] |
|  | 0.30 ± 0.04 | 10.67 ± 0.15 | 1360 ± 100 | 0.008 | 8.6 | 0.4 | Girardi et al. (2000) [29] | CCD BVI | Nilakshi (2002)[34] |
|  | 0.21 | 11.4 | 1227 ± 72 | 0.019 | 8.65 | 0.45 | Girardi et al. (2000) [29] | CCD BV | Kang et al. (2007)[35] |
|  | 0.29 | 10.6 | 1318 | 0.02 | 8.65 | 0.45 | Schaller (1992)[36] | UBV-Photoelectric | Mermilliod et al. (1996)[37] |
|  |  |  | 1502 ± 150 |  |  |  |  | Gaia DR2 parallax | Cantat-Gaudin et al. (2018)[7] |
|  |  |  | (w = 0.666 ± 0.068 mas) |  |  |  |  |  |  |

The steep core MFs of NGC 1245 ($\chi_{core} = -1.24$) is due to $[t_{rlx}(core), \tau_{core}] = [15~\text{Myr}, 103]$. In the sense its core indicates an advanced dynamical evolution stage. Its halo and overall MFs are flat ($\chi_{halo} = +0.78$, $\chi_{overall} = -0.95$) (Table 6). These MFs present signs of small-scale mass segregation in the outer region, due to the $[t_{rlx}(overall), \tau_{overall}] = [302~\text{Myr}, 5]$. In this sense, NGC 1245 does not lose too many stars from its skirt, and thus it does not seem to expose much to distorting effects such as tidal effects because of its location ($R_{GC}, \ell$) = (10.99 kpc, 146°.68). Its relatively large mass (8700 $m_\odot$) plus its large cluster dimensions ($R_{RDP}, R_{core}$) = (20.96, 2.90) pc might explain its longevity (~ 1.50 Gyr) in the Galaxy.

The core of NGC 2099 has very steep negative MF slope ($\chi_{core} = -2.67$). Its halo has also steep positive MF slope ($\chi_{halo} = +1.41$), which is consistent with Salpeter (1955)[38] ($\chi = +1.35$) and Kroupa (2001)[39] ($\chi = +1.30$) MFs. Its flat overall MF ($\chi_{overall} = +0.91$) and its $\tau_{overall} = 8$ present a sign of mass segregation. Its core and overall MFs imply that large mass stars sink in its core, whereas low-mass stars pass into its halo. Its large core evolutionary parameter, $\tau_{core} = 288$ suggests the advanced dynamical stage. Its large cluster dimensions with ($R_{RDP}, R_{core}$) = (13.07, 1.46) pc indicate an expanding core. With its location ($R_{GC}, \ell$ = (9.59 kpc, 177°.63) plus the relatively large mass, 5660 $m_\odot$, it loses its stars at low rates although it exposes to external perturbations such as tidal effects and shock waves. These evidences of NGC 2099 imply its longevity (~ 0.90 Gyr) in the Galaxy.

**Table 6.** Comparison of the literature for structural parameters (Cols.2-3), MFs (Cols.4-6) and masses (Col.7).

| Cluster | Rcore (pc) | RRDP (pc) | Xcore | Xhalo | Xoverall | Mtot(m) | literature |
|---|---|---|---|---|---|---|---|
| NGC 1245 | 2.89 ± 0.67 | 20.97 ± 1.95 | -1.24 ± 0.85 | 0.78 ± 0.72 | -0.95 ± 0.66 | 8710 ± 4640 | This paper |
|  | 3.49 ± 0.30 | 17.27 ± 1.68 | 0.32 ± 0.43 | 2.67 ± 0.64 | 1.53 ± 0.37 | 3608 ± 655 | Bukowiecki et al. (2011)[22] |
|  | - | - | - | - | -1.29 | - | Lee et al. (2012)[30] |
|  | 2.57 | - | - | - | -3.12 ± 0.27 | 1300 | Burke et al. (2004)[24] |
|  | 2.7 | - | - | - | 1.85 ± 0.30 | - | Subramaniam (2003)[31] |
|  | - | - | - | - | 1.35 ± 0.10 | - | Carraro and Patat (1994)[26] |
| NGC 2099 | 2.20 ± 0.30 | 13.07 ± 0.58 | -2.67 ± 0.47 | 1.41 ± 0.47 | 0.91 ± 0.43 | 5660 ± 2140 | This paper |
|  | 1.87 ± 0.16 | 12.97 ± 1.17 | 0.44 ± 0.17 | 2.58 ± 0.52 | 1.90 ± 0.39 | 6084 ± 1134 | Bukowiecki et al. (2011)[22] |
|  | - | - | - | - | 0.6 | 2515 | Kalirai et al. (2001)[33] |
|  | - | - | 0.22 | - | -0.67 | - | Nilakshi (2002)[34] |

From Table 6, our $\chi_{core}$ values for our target OCs are quite steep, as compared to the value of Bukowiecki et al. (2011)[22]. Our overall MFs are considerably flatter than those of Bukowiecki et al. (2011)[22]. This is the case for halo MFs. Our $\chi_{overall}$ value for NGC 1245 is quite flat, as compared to the literature values (Table 6). Note that Burke et al. (2004)[24] give $\chi_{overall} = -3.12$ down to 0.85 $m_{\odot}$. For NGC 2099, our $\chi_{overall}$ value agrees with Kalirai et al. (2001)[33]. The mean value of the mass for NGC 1245 is larger than the literature values. For NGC 2099, our mass (5660 $m_{\odot}$) is in reasonable concordance with 6084 $m_{\odot}$ of Bukowiecki et al. (2011)[22] but larger than 2515 $m_{\odot}$ of Kalirai et al. (2001)[33].

Negative $\chi$ values of the cores of the two OCs also show that the massive stars tend to be concentrated in their cores. Their large cluster and core radii (Table 6) confirm that the larger clusters tend to have larger cores, as is seen from Figure 13(a) of Camargo et al. (2010)[40]. This fact may be partly associated with large-scale mass segregation or related to initial conditions. Their large core and cluster radii also indicate an expanded core, which may suggest the presence of stellar mass black-holes. Mackey et al. (2008)[41] argue that some clusters show the expanded cores due to stellar mass black holes. Mackey and Gilmore (2003)[42] and Mackey et al. (2008)[41] also discuss that the expanded cores are the cause of growth of the limiting radii. The cluster dimensions of our sample OCs (Table 6) are consistent with the arguments of these authors.

**ACKNOWLEDGEMENTS**

We thank C. Bonatto for the algorithm for obtaining the masses. This paper has made use of results from the European Space Agency (ESA) space mission Gaia, the data from which were processed by the Gaia Data Processing and Analysis Consortium (DPAC). Funding for the DPAC has been provided by national institutions, in particular the institutions participating in the Gaia Multilateral Agreement. The Gaia mission website is http://www.cosmos.esa.int/gaia. This paper also makes use of data products from the Two Micron All Sky Survey, which is a joint project of the University of Massachusetts and the Infrared Processing and Analysis Centre/California Institute of Technology, funded by the National Aeronautics and Space Administration and the National Science Foundation. This research has made use of the WEBDA database, operated at the Institute for Astronomy of the University of Vienna.